\documentstyle[12pt]{article}

\begin{document}

\begin{center}

{\large \bf Gluon Confinement } \\

\vspace{2.0cm}

M. Novello and V. A. De Lorenci \\
\vspace{0.1cm}
{\it Centro Brasileiro de Pesquisas F\'{\i}sicas,} \\
\vspace{0.1cm}
{\it Rua Dr. Xavier Sigaud, 150, Urca} \\
\vspace{0.1cm}
{\it 22290-180 -- Rio de Janeiro -- RJ. Brazil.} \\
\vspace{0.1cm}
and \\
\vspace{0.1cm}

E. Elbaz \\
\vspace{0.1cm}
{\it Institut de Physique Nucl\'eaire de Lyon IN2P3-CNRS} \\
\vspace{0.1cm}
{\it Universit\'e Claude Bernard} \\
\vspace{0.1cm}
{\it 43 Bd du 11 Novembre 1918, F-69622 Villeurbanne Cedex, France.}

\vspace{2cm}

\begin{abstract}
In this paper we present a new model for a gauge field theory such that 
self-interacting spin-one particles can be confined in a compact 
domain. The necessary conditions to produce the confinement 
appear already in the properties of the eikonal structure 
generated by the particular choice of the dynamics. The gluons do not
follow along Minkowski null cone and thus can be equivalently interpreted as 
if they were massive.
\end{abstract}

\vspace{2cm}

\end{center}

Pacs number: 11.15-q, 12.38.Aw.

\newpage

It is a common knowledge that the necessary conditions for a field theory 
to produce confinement is to be non-linear. Although 
this seems to be a good requirement, it is not enough. Since such a 
confinement, for instance, in the SU(3) non-Abelian gauge theory 
containing an octet of massless vector gluons, is still not 
available \cite{Pre}, a 
natural question appears: is there something missing in the 
Yang-Mills theory? The aim of this letter is to exhibit in a simple 
version a pattern that answers affirmatively to this question.  

In its classical version, the confinement of massless spin-one particles 
can be interpreted as the deformation of their surfaces of propagation, the 
corresponding {\it light cones}, in such a way that the gluons 
encounter an unsurmountable barrier that forbids 
them to get outside the confinement region. In other terms, it appears, 
for the external world, as a situation that can be described equivalently 
in terms of the formation of a horizon. Such a deformation does not 
occur in the Minkowskian structure of spacetime. 
How to create a similar effect
in a scenario containing just a set of spin-one self-interacting fields? 

To obtain the required property that the information carried on by the 
gluons operate in a different way than the Minkowski null cones, 
the first step is to change the dynamics. 
The reason for this is related to the fact that massless Yang-Mills 
(YM) particles travel along the Minkowski null cone\footnote{The need for 
this modification comes from the property of the structure 
of geodesics in Minkowski space, that imposes that any 
particle that follows null cones cannot be bounded in a compact region.}. 
This property is the same as in the Abelian case and has its origin 
as a direct consequence on the construction of the YM model that makes 
all non-linearity of this theory to be restricted uniquely to the 
algebraical dependence of 
the field $F^{a}_{\mu\nu}$  on the potential $A_{\mu}^{a}$. 
To change this in the realm of the gluon interaction, conserving the 
colour covariance, one should modify conveniently the Lagrangian. 
This is precisely the case which we will consider here 
and constitutes the basis for the actual new ingredient of our model. 
We shall see that a slight modification of the YM theory\footnote{We remark 
that the major part, and by far the most important one, of 
Yang-Mills theory is maintained.} concerning its Lagrangian seems to be the 
key to the understanding of the confinement problem. 

We start by considering a set $A_{\mu}^{a}$ of colour multiplet 
that constitutes a Yang-Mills field of a non-Abelian 
theory\footnote{We have in mind, for instance, 
the standard $SU(3)$ non-Abelian QCD model.}
with $F_{\mu\nu}^{a}$ as the corresponding field. Let ${\bf F}$ be the 
invariant under space-time\footnote{The Minkowski metric is $\eta_{\mu\nu} =
diag(+,-,-,-)$.} and internal colour coordinates, defined by 
\begin{equation}
{\bf F} \equiv \vec{F}^{\mu\nu}.\,\vec{F}_{\mu\nu} = 
F^{a\,\mu\nu}\, F_{a\,\mu\nu}.
\end{equation}
The dynamics set up in the Yang-Mills approach mimetizes Maxwell 
electrodynamics by the identification of the Lagrangian to such 
quantity, that is, $L_{\mbox{\tiny YM}} = {\bf F}$. 
Should this be taken as an irretrievable paradigm? 
Does a change on this hypothesis, in the hadron 
world, yield the desirable consequences? Before answering to this, 
let us make a small comment on the classical description.

From a broad principle the Lagrangian should have the general non-linear form
\begin{equation}
L = L({\bf F}).
\end{equation}
Although one can go further without being necessary to specify the 
form of such a functional, in order to have a definite model that exhibits 
in a simple manner the main aspects of our ideas, we limit all our 
considerations here to the specific model provided by:
\begin{equation}
L_{ \mbox{\tiny NDE}}  = \epsilon_{s}\left(1 -\, e^{\frac{{\bf F}}{4\epsilon_{s}}}\right)
\label{model}
\end{equation}
in which $\epsilon_{s}$ is a constant. 
The corresponding equation of motion is given by 
\begin{equation}
D^{ac}_{\nu}\left\{ e^{\frac{{\bf F}}{4\epsilon_{s}}}\, 
F_{c}^{\mu\nu}\right\} = 0
\label{eqmov}
\end{equation}
that is
\begin{equation}
\left[\delta_{cd}\partial_{\nu}  + g\,c_{acd}A^{a}_{\nu}\right]
\left( e^{\frac{{\bf F}}{4\epsilon_{s}}}\, F^{d\,\mu\nu}\right) = 0,
\end{equation}
where $c_{abc}$ are the constants of the structure of the gauge
group and $g$ is the strong interaction coupling constant. 
To proceed with the examination of the corresponding behavior of 
the classical gluons in such non-linear theory there is no better way than to 
consider the very high energy case through the analysis of 
the eikonal. In the standard Yang-Mills 
dynamics, the eikonal is nothing but null-cones of the Minkowski 
background spacetime, as in Maxwell theory. This is not the case for 
our Lagrangian. Indeed, there exist examples of spin-one theories in 
which the eikonal follows 
null geodesics in an effective geometry which depends not only on the 
background metric but also on the field properties. This has been
shown in the case of pure non-linear electrodynamics \cite{Plebansky}. 
This result keeps being valid in the non-Abelian gauge theory, as we
will now show\footnote{At the basis of this property 
rests the fact that the dependence of the group connection 
on the potential do not contain derivatives but only an algebraic
form. Standard conditions for the wave disturbances, like 
Hadamard\rq s structure, imply that this sector of the non-linearity does not 
affect the velocity of propagation.}.

Let $\Sigma$ be a surface of discontinuity for the gauge field. 
Following Hadamard's \cite{Hadamard} condition we take the 
potential and the field as 
being continuous through $\Sigma$ but having its first derivative 
discontinuous, that is:
\begin{equation}
[F_{\mu\nu}^{a}]_{\Sigma} = 0,
\label{gw1}
\end{equation} 
and   
\begin{equation}
[\partial_{\lambda}\,F^{a}_{\mu\nu}]_{\Sigma} = f^{a}_{\mu\nu} k_{\lambda},
\label{gw2}
\end{equation} 
in which the symbol $[ J ]_{\Sigma}$ represents the discontinuity 
of the function $J$ through the surface $\Sigma$ and $k_{\lambda}$ is
the normal to $\Sigma$.

Applying these conditions into the equation of motion (\ref{eqmov}) we obtain
\begin{equation}
f^{\mu\nu}_{a}\, k_{\nu} + \frac{1}{2\epsilon_{s}} \,\xi F^{\mu\nu}_{a}\, k_{\nu} = 0,
\label{gw3}
\end{equation} 
where $\xi$ is defined by 
\begin{equation}
\xi \equiv F^{\alpha\beta}_{a} \, f_{\alpha\beta}^{a}.
\end{equation}
From the cyclic identity,
\begin{equation}
D^{bc}_{\lambda}\,F^{a}_{\mu\nu} + D^{bc}_{\mu}\,F^{a}_{\nu\lambda}
+ D^{bc}_{\nu}\,F^{a}_{\lambda\mu} = 0
\label{ciclo}
\end{equation}
and using the above continuity conditions of the potential and the fields 
yields
\begin{equation}
f_{\mu\nu}^{a} k_{\lambda} + f_{\nu\lambda}^{a} k_{\mu} + 
f_{\lambda\mu}^{a} k_{\nu} = 0.
\label{gw4}
\end{equation}
Multiplying this equation by $k_{\lambda}\, F^{\mu\nu}_{a}$ yields
\begin{equation}
\xi k_{\nu} \,k_{\mu} \eta^{\mu\nu} + 2 
\,F^{\mu\nu}_{a}\,f^{a}_{\nu}\mbox{}^{\lambda} 
\,k_{\lambda} \, k_{\mu} = 0. 
\end{equation}
Using Eq. (\ref{gw3}) in this expression and
after some algebraic manipulations the equation of propagation of the
disturbances is obtained:
\begin{equation}
\left\{\eta^{\mu\nu} + \Lambda^{\mu\nu} \right\} k_{\mu} k_{\nu} = 0
\label{gww4}
\end{equation}
in which the quantity $\Lambda^{\mu\nu}$ is provided by
\begin{equation}
\Lambda^{\mu\nu} \equiv - \frac{1}{\epsilon_{s}} \,
F^{a\,\mu\lambda} \,F_{a\,\lambda}\mbox{}^{\nu}.
\label{lambda}
\end{equation}

The net effect of this modification of the non-linearity of the 
Yang-Mills theory can thus be summarized in the following property: 
{\it The disturbances of the gauge field controlled by the non-linear
Lagrangian $L_{\mbox{\tiny NDE}}$ 
propagate through null geodesics of the modified effective geometry 
given by:} 
\begin{equation}
g^{\mu\nu} = \eta^{\mu\nu}  - \frac{1}{\epsilon_{s}} 
\,F^{a\,\mu\lambda} \,F_{a\,\lambda}\mbox{}^{\nu}.
\label{geffec}
\end{equation}

Let us emphasize that this property stands only from the structural form of 
the dynamics of our theory. To avoid misunderstanding\footnote{The 
reason for this additional assertion is due to the fact that 
modifications on the underlying geometry are traditionally 
supposed to be connected 
to gravitational forces. We would like to stress that this {\bf is not}
the case here.} we state:
\begin{itemize}
 \item{{\bf This geometry modification is a pure spin-one non-linear 
phenomenon.}}
\end{itemize}

Thus, we conclude from the above statement that gluon dynamics 
can be examined through the properties of 
null geodesics in the modified geometry. The fact that in this theory 
massless spin-one particles do not follow Minkowski null cone 
occurs as a direct consequence of the particular 
non-linear dynamics used in our model, which is distinct from the one 
contained in standard Yang-Mills theory.

In order to show the confinement induced by such non-linear model
there is no better way than to investigate the behavior of the
null geodesics in this geometry. 
For our purposes, we restrict ourselves here on the analysis 
of a spherically symmetric and static solution. A direct computation 
shows that in the spherical coordinate system such a 
particular solution can be found uniquely in terms of a radial 
component given by:
\begin{eqnarray}
F^{a}_{01} &=& f(r)\,n^{a}, 
\end{eqnarray}
in which $n^{a}$ is a constant vector in the colour space and 
$f(r)$ is given by the relation:
\begin{equation}
e^{-\,\frac{f^2}{2\epsilon_{s}}} \,f = \frac{Q}{r^{2}}.
\label{relationf}
\end{equation}
The parameter $Q$ is related to distribution of the charge $q(r)$:
\begin{equation}
Q = \int {\rm d}^3x q(r).
\end{equation}

From the standard definition of the energy momentum tensor we obtain:
\begin{equation}
T_{\mu\nu} = - \epsilon_{s}\left(1 -\, e^{\frac{{\bf F}}{4\epsilon_{s}}} 
\right) \eta_{\mu\nu} 
+ e^{\frac{{\bf F}}{4\epsilon_{s}}} \,\vec{F}_{\mu\alpha} . 
\vec{F}^{\alpha}\mbox{}_{\nu},
\end{equation}
and for the density of energy $T_{00}$ results:
\begin{equation}
T^{0}\mbox{}_{0} = \frac{Q}{r^2}\left[\frac{\epsilon_{s} + f^2}{f} 
- \epsilon_{s}\right].
\end{equation}

A remarkable  characteristic of the geometry induced by this non linear 
spin-1 field theory may be made explicit by looking into the line element
of the corresponding effective geometry in which the massless spin-1 
particles travel. From Eq. (\ref{geffec}) and the
spherically symmetric solution we obtain\footnote{Note that in order
to write Eq. (\ref{lineelement}), that is,  
${\rm d}s^2 = g_{\mu\nu}{\rm d}x^{\mu}{\rm d}x^{\nu}$ we need to know
the inverse metric $g_{\mu\nu}$ defined by
$g_{\mu\nu}g^{\mu\rho} = \delta^{\rho}_{\nu}$}:
\begin{equation}
{\rm d}s^2 = \left(1 - \frac{f^2}{\epsilon_{s}}\right)^{-1} {\rm d}t^2 -  
\left(1 - \frac{f^2}{\epsilon_{s}}\right)^{-1} 
{\rm d}r^2 - r^2\,{\rm d}\theta^2 -  r^2\,\sin^2 \theta \,
{\rm d}\varphi^2,
\label{lineelement}
\end{equation}
in which $f(r)$ is given by Eq. (\ref{relationf}).
A direct inspection of the line element shows that there is a value of the 
function $f(r)$ in which the $g_{00}$ and $g_{11}$ metric components 
are singular. This point corresponds to the solution
\begin{equation}
f(r) = \sqrt{\epsilon_{s}}.
\label{critical}
\end{equation}
Note however that this is not a true physical singularity. Indeed, 
let us look into
the effective potential in order to prove this\footnote{It
seems worthwhile to quote here an analogous situation occuring in Einstein's
theory of general relativity in the case of Schwarzschild solution. 
In both situations we are dealing with a horizon and not with true 
singularity.}. 

To obtain the form of the potential it is enough to look for the 
radial equation of motion of the geodesics in this solution, 
as a function of the proper time. The simplest way to arrive at this 
is by means of the variational principle
\begin{equation}
\delta \,\int\left[ \left(1 - 
\frac{f^2}{\epsilon_{s}}\right)^{-1} \dot{t}^2 -  
\left(1 - \frac{f^2}{\epsilon_{s}}\right)^{-1} 
\dot{r}^2 - r^2\,\dot{\theta}^2 -  r^2\,\sin^2 \theta \,\dot{\varphi}^2 \right]
ds = 0
\end{equation}
in which we have used the effective geometry --- as it appears in Eq. 
(\ref{lineelement}). A dot means proper time derivative.
The radial dependence yields:
\begin{equation} 
\dot{r}^2 + V_{eff} = l_{0}^2
\end{equation}
in which the potential $V_{eff}$ has the form:
\begin{equation} 
V_{eff} = \frac{\left[\epsilon_{s}-f(r)^2\right]h_{0}^{2}}{\epsilon_{s}r^2} 
- \frac{\left[\epsilon_{s}-f(r)^2\right]^{2}l_{0}^{2}}{\epsilon_{s}^{2}} 
+ l_{0}^{2},
\label{potential}
\end{equation}
and $h_{0}$ and $l_{0}$ are constants of motion.

The form of this potential shows that 
the gluons in the $L_{\mbox{\tiny NDE}}$ non-linear theory behave as 
particles endowed with energy $l_{0}^2$, immersed in a 
central field of forces characterized by the potential $V_{eff}$. 
 
Let us summarize what we have achieved. Massless spin-one particles (gluons)
obeying Yang-Mills dynamics travel along null cones. 
In a Minkowski spacetime it is a hard task to confine such particles in a 
compact region, once this phenomenon should be associated with the presence 
of a singular horizon. We are then led to a modification of the self 
interaction properties of the gluons. We present here a toy model that 
can be equivalently described in terms of an 
effective change of the background geometry\footnote{This consequence
of the paths of the gluons not follows along Minkowski null cone
can be interpreted in an equivalent way, as if the gluons had a mass.}. 
We analyse a particular example of a static, spherically symmetric solution
and proceed to the exam of the corresponding null geodesics, the 
gluon paths, in the associated geometry. It then follows that the 
behavior of gluons can be examined in terms of the metric structure
(\ref{lineelement}) and the potential given 
in Eq. (\ref{potential}) showing, through the appearance of a
horizon, the required confining feature. This 
result allows us to argue that the final solution
of the confinement of the gluons could be found along these lines.

\section*{Acknowledgements}
We would like to thank Dr. I. Bediaga for calling our attention
to the confinement problem. This work was supported by Conselho Nacional de Desenvolvimento Cient\'{\i}fico e Tecnol\'ogico (CNPq) of Brazil.

\section*{Appendix: Dielectric Constant and the Effective Geometry}

In order to show the non-familiar reader the treatment that involves  
dealing with the propagation of the non-linear theory as a modification 
of the background geometry, we will present here the simplest possible case 
of the standard Maxwell theory in a dielectric medium. We will show how 
it is possible to present the wave propagation of linear electrodynamics in 
a medium in terms of a modified geometry of the spacetime.

In this appendix we take the Maxwell theory in a medium such that the 
electromagnetic field is represented by two anti-symmetric tensors 
$F_{\mu\nu}$ and $P_{\mu\nu}$ given in terms of the electric and 
magnetic vectors, as seen by an arbitrary observer 
endowed with a four-velocity $v^{\mu}$, by the standard expressions:
\begin{equation}
F_{\mu\nu} = E_{\mu}\, v_{\nu} - E_{\nu}\, v_{\mu} + 
\eta^{\rho\sigma}_{\mu\nu} \, v_{\rho} \, H_{\sigma}
\end{equation}
and
\begin{equation}
P_{\mu\nu} = D_{\mu}\, v_{\nu} - D_{\nu}\, v_{\mu} + 
\eta^{\rho\sigma}_{\mu\nu} \, v_{\rho} \, B_{\sigma}.
\end{equation}
Maxwell equations are:
\begin{equation}
\partial^{\nu} \,F^{*}_{\mu\nu} = 0
\end{equation}
\begin{equation}
\partial^{\nu} \,P_{\mu\nu} = 0.
\end{equation}
Following Hadamard, we consider the discontinuities on the fields 
as given by
\begin{eqnarray}
\left[\partial_{\lambda}\, E_{\mu\nu}\right]_{\Sigma} &=& k_{\lambda}\, 
e_{\mu\nu}
\nonumber\\
\left[\partial_{\lambda}\, D_{\mu\nu}\right]_{\Sigma} &=& k_{\lambda}\, 
d_{\mu\nu}
\nonumber\\
\left[\partial_{\lambda}\, H_{\mu\nu}\right]_{\Sigma} &=& k_{\lambda}\, 
h_{\mu\nu}
\nonumber\\
\left[\partial_{\lambda}\, B_{\mu\nu}\right]_{\Sigma} &=& k_{\lambda}\, 
b_{\mu\nu}.
\end{eqnarray} 
Using the constitutive relations\footnote{We deal here with the
simplest case of linear isotropic relations, just for didactic reasons.}
\begin{eqnarray}
d_{\mu} &=&  \epsilon \, e_{\mu} 
\\
b_{\mu} &=&   \frac{h_{\mu}}{\mu} 
\end{eqnarray}
one obtains after a straightforward calculation
\begin{equation}
k_{\mu}\, k_{\nu} \left( \eta^{\mu\nu} + (\epsilon\, \mu - 1) v^{\mu}
\,v^{\nu} \right) = 0.
\end{equation}
This shows that even the simple case of the evolution of the 
wave front in standard Maxwell equation in a medium can be
interpreted in terms of an effective geometry $g^{\mu\nu}$ that
depends not only on the medium properties $\epsilon$ and $\mu$, but 
also on the observer\rq s velocity, given by: 
\begin{equation}
g^{\mu\nu} \equiv  \eta^{\mu\nu} + (\epsilon\, \mu - 1) v^{\mu}
\,v^{\nu}.
\end{equation}
This ends our proof.

\end{document}